\documentclass{PoS}

\title{$\psi$(2S) production and nuclear modification factor in nucleus--nucleus collisions with ALICE}

\ShortTitle{$\psi$(2S) production and nuclear modification factor in nucleus--nucleus collisions}

\author{\speaker{Biswarup Paul} %
         \thanks{on behalf of the ALICE Collaboration}\\
       \\ University and INFN Cagliari,\\
          Cittadella Universitaria di Monserrato, 09042, Monserrato (CA), Italy\\
          E-mail: \email{biswarup.paul@cern.ch}}

\abstract{
Charmonium production is a probe sensitive to deconfinement in nucleus--nucleus collisions. The production of J/$\psi$ via regeneration within the QGP or at the phase boundary has been identified as an important ingredient for the description of the observed centrality and $p_{\rm T}$ dependence at the LHC. $\psi$(2S) production relative to J/$\psi$ is one possible discriminator between the two different regeneration scenarios. At RHIC and at the LHC, there is so far no significant observation of the $\psi$(2S) in nucleus--nucleus collisions in central events at low transverse momentum, where regeneration is the dominating process.  The combined Run 2 data set of ALICE allows to extract a significant $\psi$(2S) signal in such a kinematic region at forward rapidity in the dimuon decay channel.  In this contribution, we present for the first time results on the $\psi$(2S)-to-J/$\psi$ double ratio and the $\psi$(2S) nuclear modification factor  in Pb--Pb collisions at $\sqrt{s_{\rm NN}} = 5.02$ TeV, calculated with respect to a new pp reference with improved precision. Results are compared with model calculations.

}

\FullConference{%
  41st International Conference on High Energy physics - ICHEP2022\\
  6-13 July, 2022\\
  Bologna, Italy
}

\begin{document}

\section{Introduction}
At the extreme temperatures and energy densities produced in ultrarelativistic collisions of heavy nuclei, hadronic matter undergoes a transition into a state of deconfined quarks and gluons over distances much larger than the hadronic size ($\sim$ 1 fm), known as quark-gluon plasma (QGP). Charmonium (vector meson consisting of charm quark and anti-charm quark) is expected to be dissociated in a QGP by color screening and hence it is used as one of the most prominent probe to investigate the properties of the QGP~\cite{satz,satz2}. Differences in the binding energies lead to a sequential melting of the charmonium states with increasing temperature of the QGP. Because of the larger size (by a factor 2) and weaker binding energy (by more than a factor 10) of the $\psi$(2S) state compared to J/$\psi$, $\psi$(2S) is expected to be strongly suppressed that J/$\psi$. At LHC energies, due to the large increase of the $c\overline c$ production cross-section with the collision energy, there is a possibility of charmonium production via recombination of ${c}$ and $\overline {c}$. Thus, the observation of charmonium production in nucleus-nucleus collisions via recombination also constitutes an evidence of QGP formation. $\psi$(2S) production relative to J/$\psi$ represents one possible discriminator between the two different regeneration scenarios, regeneration within the QGP~\cite{reg_QGP} or regeneration at the phase boundary~\cite{reg_phase, reg_phase2}. The $\psi(2\rm S)$-to-J/$\psi$ cross-section ratio is predicted to be very sensitive to the details of the recombination mechanism. Experimentally this ratio is interesting as most of the systematic uncertainties cancel, with the remaining systematic uncertainties being only due to the signal extraction and the efficiency evaluation. On the theory side, this ratio is also weakly dependent on the total charm production cross section employed as inputs to the models. Thus, the two effects, suppression and recombination, act in opposite directions and the comparison of the different charmonium states can provide insights to the evolution of the relative contributions of the two processes. The pp results~\cite{alice_pp,bpaul} for the charmonium provide a baseline for the nuclear modification factor of charmonium production in Pb--Pb collisions.

\section{ALICE detector and data samples}
The ALICE experiment has studied inclusive $\psi$(2S) production in Pb--Pb collisions at $\sqrt{s}_{\rm NN} = \mbox{5.02 TeV}$ through its dimuon decay channel. Muons are identified and tracked in the Muon Spectrometer, which covers the pseudorapidity range $-4<\eta<-2.5$~\cite{alice}. The pixel layers of the Inner Tracking System (ITS) allow the vertex determination, while forward VZERO scintillators are used for triggering purposes. The VZERO is also used to determine the centrality of the collisions. The data samples used in this analysis, were collected in 2015 and 2018, correspond to an integrated luminosity $L_{\rm int} \sim 750~ \mu{\rm b}^{-1}$. The data sample of pp collisions at $\sqrt{s}_{\rm NN} = \mbox{5.02 TeV}$ was collected in 2017 and corresponds to an integrated luminosity $L_{\rm int} \sim 1230~ {\rm nb}^{-1}$.

\section{Results}
\subsection{pp results}
\begin{figure}[ht]
\centering
\includegraphics[scale=0.37]{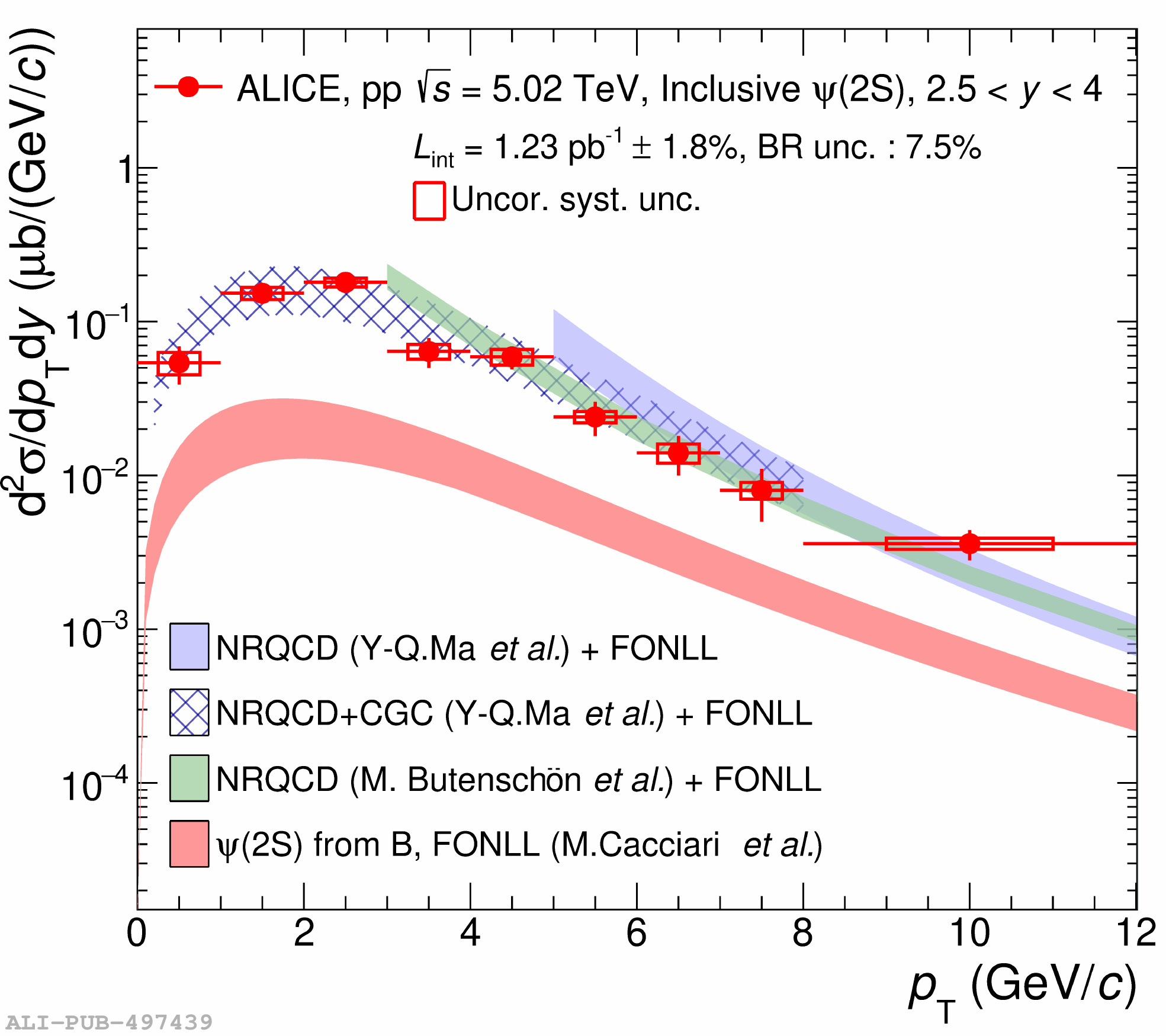}
\includegraphics[scale=0.37]{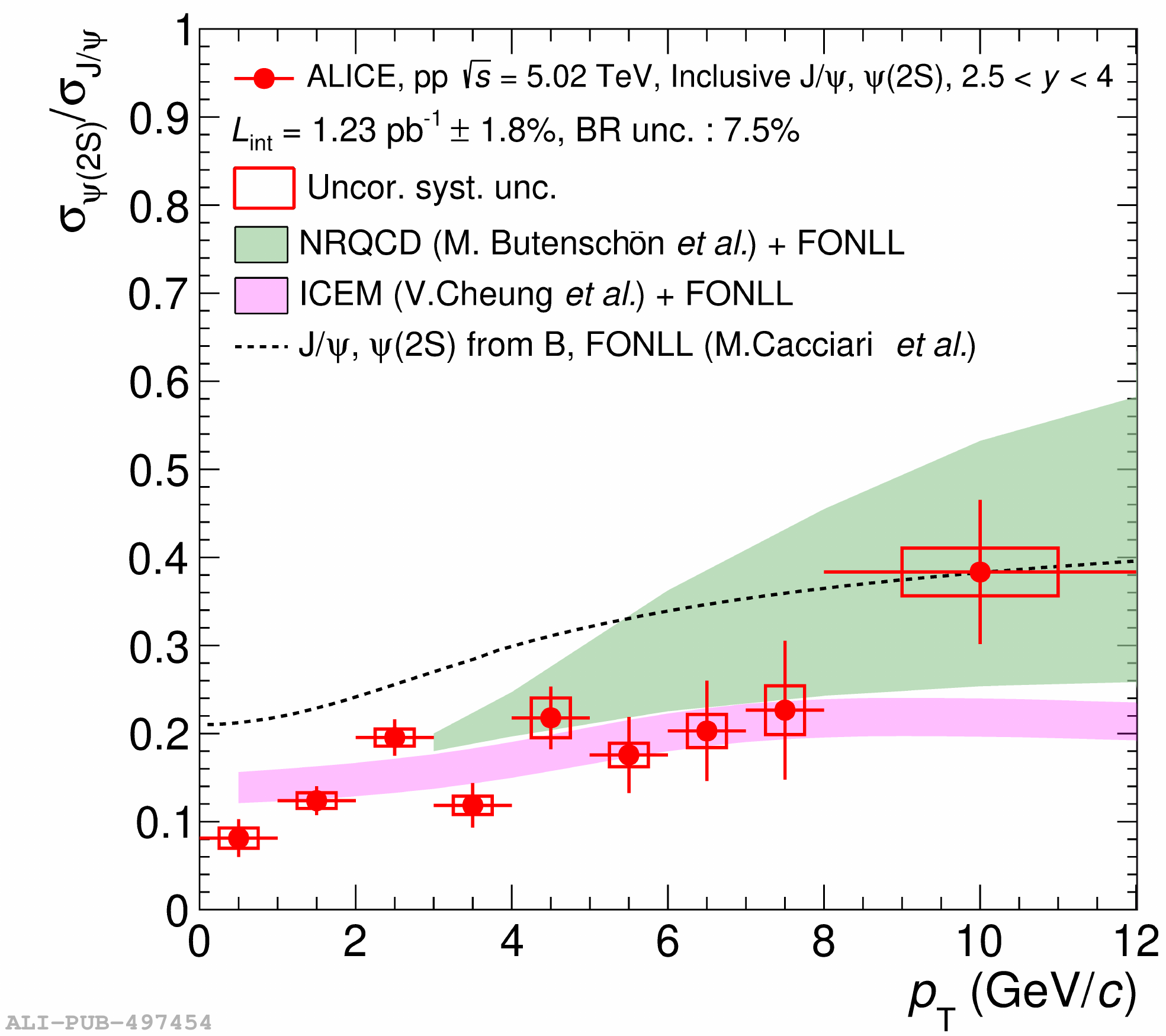}
\vspace{-0.2cm}
\caption{\label{Psi2Sppref}The left panel shows the $p_{\rm T}$ dependence for the inclusive $\psi$(2S) production cross section in pp collisions at $\sqrt{s}$ = 5.02 TeV. The results are compared with the theory predictions based on NRQCD~\cite{But,QMa,QMa2} models. The calculation of the non-prompt contribution from FONLL calculations~\cite{MCacci} are also shown separately. The right panel shows the inclusive $\psi$(2S)-to-J/$\psi$ cross section ratio as a function of $p_{\rm T}$ compared with theoretical calculations~\cite{MCacci,But,Cheung}.}
\end{figure}

The inclusive $\psi$(2S) production cross section in pp collisions at $\sqrt{s}$ = 5.02 TeV at forward rapidity as a function of $p_{\rm T}$ is shown in the left panel of Fig.~\ref{Psi2Sppref}. An improvement of a factor $\sim$ 3 for the statistical uncertainty is obtained for the most recent data set compared to the previous publication~\cite{Psi2S_Old}. Thanks to the large statistics, the first results on the $p_{\rm T}$ and $y$ dependence of the inclusive $\psi$(2S) cross section for 2.5 $ < y <$ 4 in pp collisions at $\sqrt{s}$ = 5.02 TeV are obtained~\cite{alice_pp}. The result is also compared with theoretical models. The non-prompt $\psi$(2S) contribution from FONLL~\cite{MCacci} is also shown in Fig.~\ref{Psi2Sppref} and it is summed to all theoretical predictions. The NRQCD calculation from Butensch$\ddot{\rm o}$n et al.~\cite{But} agrees with the experimental data for 4 $< p_{\rm T} <$ 12 GeV/$c$, and the NRQCD calculation from Ma et al.~\cite{QMa} describes well the data except for 5 $< p_{\rm T} < 6$ GeV/$c$, where it overpredicts them. The NRQCD+CGC~\cite{QMa2} model provide a good description of the $\psi$(2S) cross section as a function of $p_{\rm T}$. 

The $p_{\rm T}$ dependence of the $\psi$(2S)-to-J/$\psi$ cross section ratio is shown in the right panel of Fig.~\ref{Psi2Sppref}. The boxes represent the uncorrelated systematic uncertainties due to the MC input shapes and the signal extraction. The branching-ratio uncertainties, fully correlated versus $p_{\rm T}$, is reported in the legend of Fig.~\ref{Psi2Sppref}. All the other systematic uncertainties are correlated over the two resonances and cancel out in the ratio.  As in previous sections, the non-prompt contribution from FONLL~\cite{MCacci} is added to all theoretical calculations. The NRQCD calculations from Butensch$\ddot{\rm o}$n et al.[43] describe well the $p_{\rm T}$ dependence of the cross section ratio within the large model uncertainties. A good description of the trend of the $\psi$(2S)-to-J/$\psi$ cross section ratio as a function of $p_{\rm T}$ is also provided by the ICEM model~\cite{Cheung}. This inclusive $\psi$(2S) production cross section and $\psi$(2S)-to-J/$\psi$ cross section ratio in pp collisions have been used as a baseline to calculate $\psi$(2S) $R_{\rm AA}$ and $\psi$(2S)-to-J/$\psi$ double ratio in Pb--Pb collisions, discussed below.

\subsection{Pb--Pb results}

\begin{figure}[ht]
\centering
\includegraphics[scale=0.37]{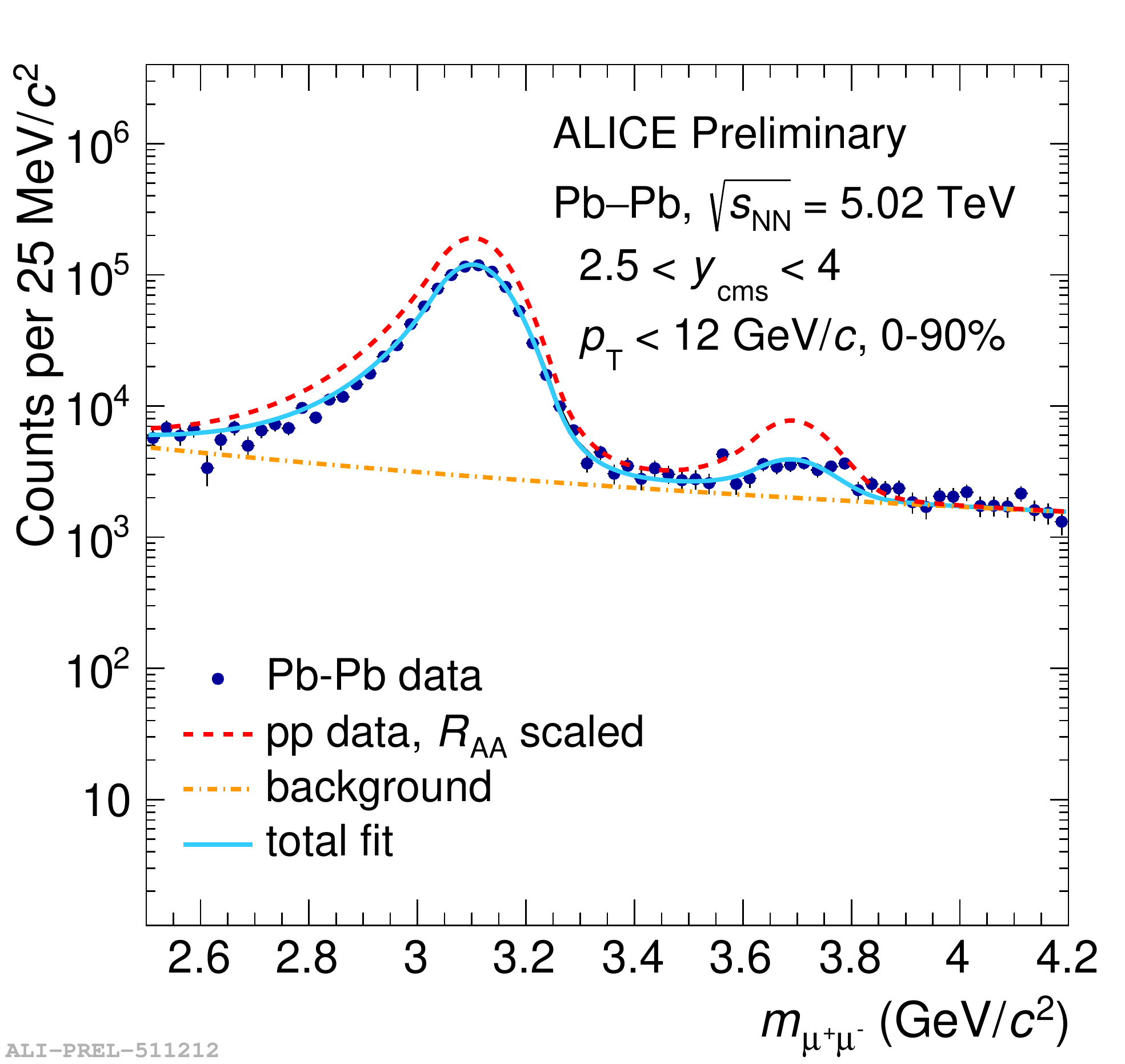}
\vspace{-0.2cm}
\caption{\label{Invmass}Fit of the $p_{\rm T}$-integrated opposite-sign dimuon invariant-mass spectrum for 0--90\% centrality at forward rapidity in Pb--Pb collisions at $\sqrt{s}_{\rm NN} = \mbox{5.02 TeV}$. The contributions of J/$\psi$, $\psi$(2S) and background continuum are shown, as well as the result of the fit. The dashed line represents the expected yields in case collision scaling holds, i.e., $R_{\rm AA}$ = 1.}
\end{figure}

Fig.~\ref{Invmass} shows example of fit to the opposite-sign dimuon dimuon invariant-mass distributions with background subtraction using the event-mixing technique for 0--90\% centrality and 0 $< p_{\rm T} <$ 12 GeV/$c$ at forward rapirity (2.5 $< y <$ 4) in Pb--Pb collisions at $\sqrt{s}_{\rm NN} = \mbox{5.02 TeV}$. The contributions of J/$\psi$, $\psi$(2S) and background continuum are shown, as well as the result of the fit. The dashed line represents the expected yields in case collision scaling holds, i.e., $R_{\rm AA}$ = 1.

\begin{figure}[ht]
\centering
\includegraphics[scale=0.37]{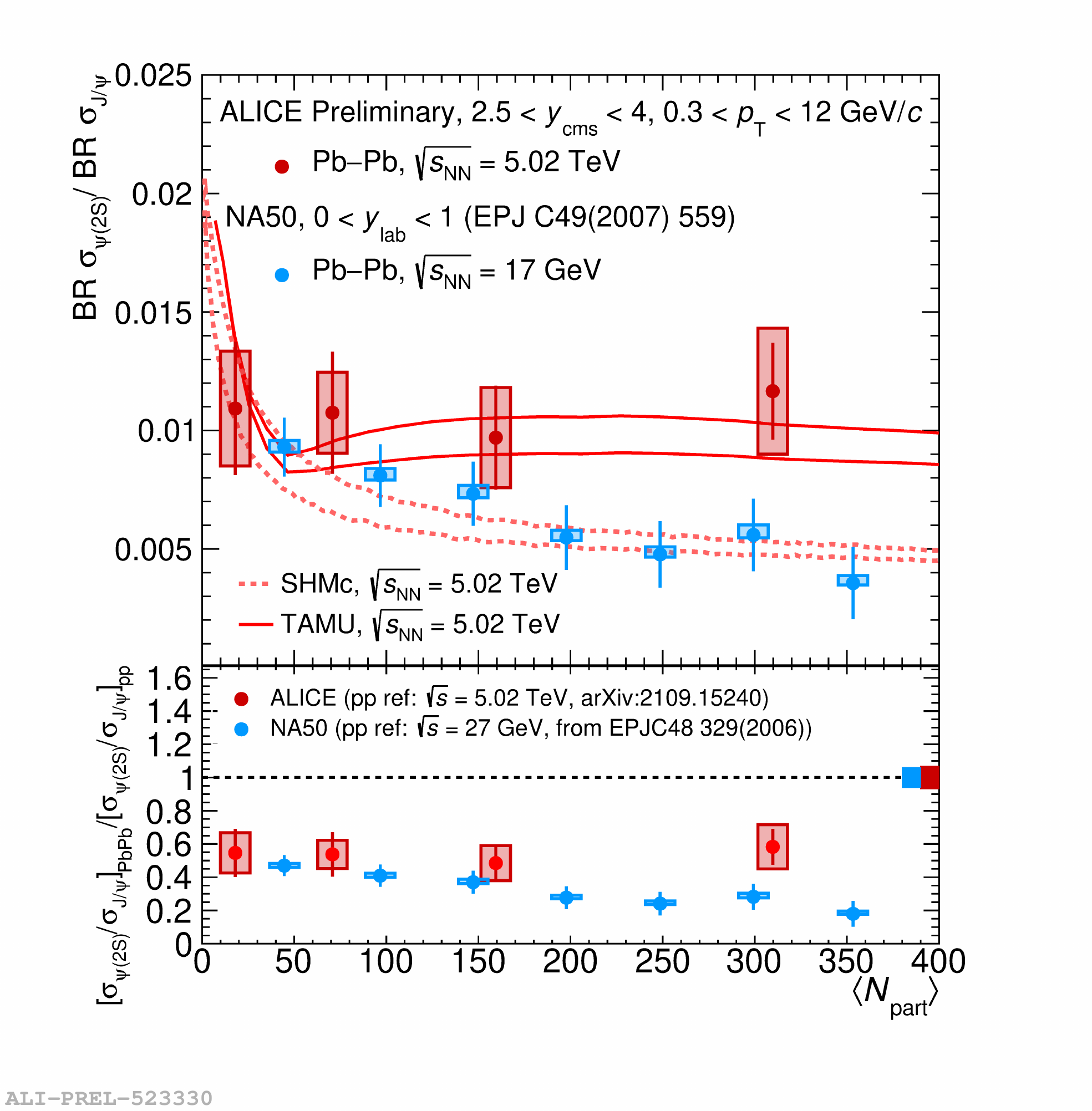}
\includegraphics[scale=0.37]{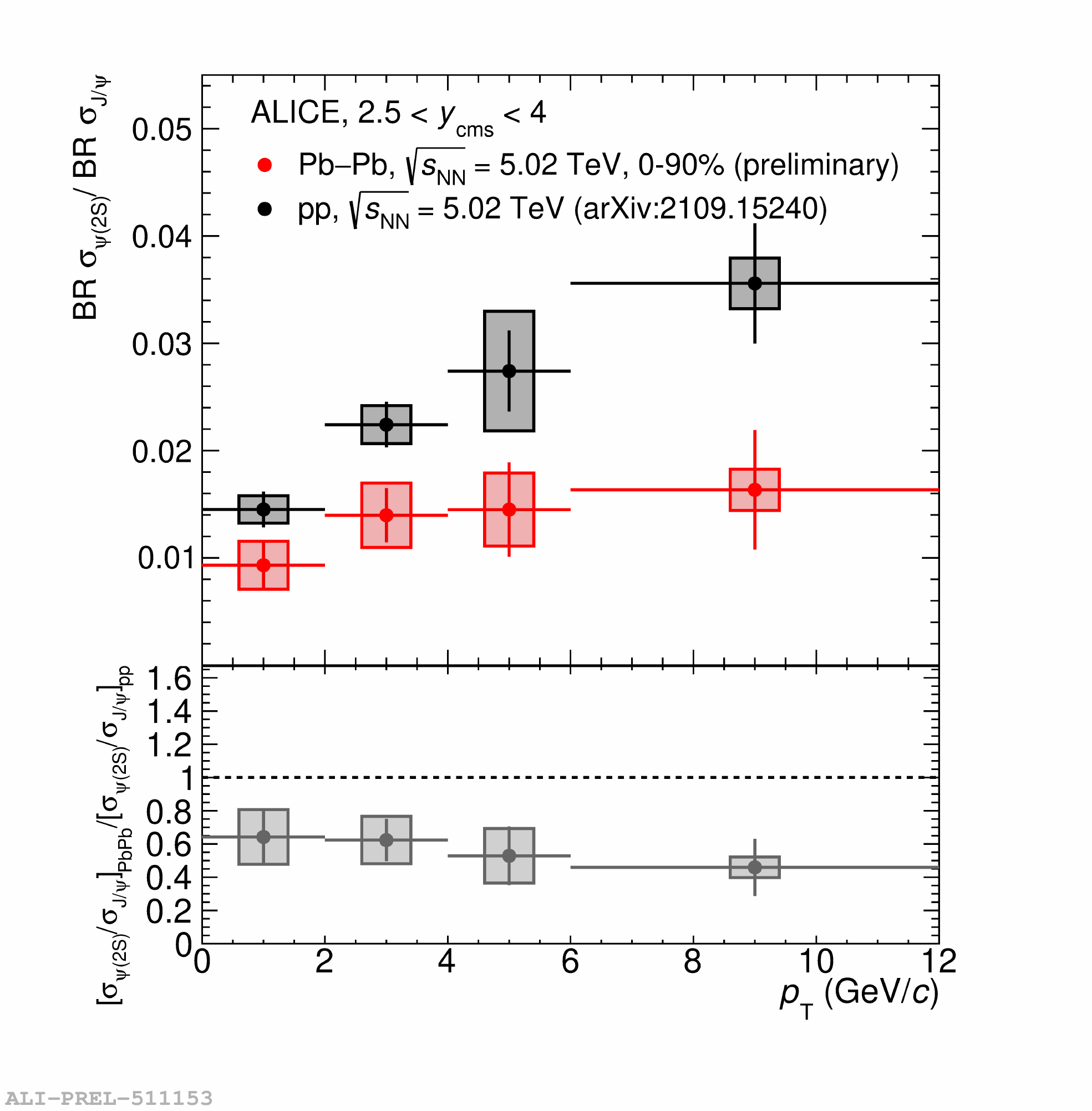}
\vspace{-0.2cm}
\caption{\label{Psi2SJPsiratios}$\psi$(2S)-to-J/$\psi$ cross section ratio measured by the ALICE collaboration in Pb--Pb collisions at $\sqrt{s_{\rm NN}}$ = 5.02 TeV as a function of $\langle N_{\rm part}\rangle$ (left) and $p_{\rm T}$ (right). In the left panel, NA50 measurements at SPS carried out at $\sqrt{s_{\rm NN}}$ = 17 GeV~\cite{NA50} are also shown. The results, are compared with theoretical predictions from TAMU~\cite{TAMU} and SHMc~\cite{SHMc,SHMc2}. Bottom panels show the $\psi$(2S)-to-J/$\psi$ ratio normalized to the corresponding pp value (double ratio).}
\end{figure}

The $\psi$(2S)-to-J/$\psi$ cross section ratio (not corrected for the branching ratios of the dimuon decay) measured by the ALICE collaboration in Pb--Pb collisions at $\sqrt{s}_{\rm NN} = \mbox{5.02 TeV}$ at forward rapidity as function of centrality, expressed in terms of average number of participant nucleons $\langle N_{\rm part}\rangle$, is shown in left panel of Fig.~\ref{Psi2SJPsiratios}. The $\psi$(2S)-to-J/$\psi$ double ratio is shown in the bottom panel of Fig.~\ref{Psi2SJPsiratios}, indicating a $\psi$(2S) suppression effect by 50\% in Pb--Pb with respect to pp collisions. Flat centrality dependence is observed within uncertainties. The centrality dependence of both the ratios are compared with NA50 results in Pb--Pb collisions in 0 $< y_{\rm Lab} < 1$ at $\sqrt{s}_{\rm NN}$ = 17 GeV~\cite{NA50} and NA50 exhibit a stronger centrality dependence, reaching smaller values in central collisions. TAMU model~\cite{TAMU} reproduces the centrality dependence of $\psi$(2S)-to-J/$\psi$ ratio, while SHMc~\cite{SHMc,SHMc2} tends to underestimate the result in central Pb--Pb collisions.

In the right panel of Fig.~\ref{Psi2SJPsiratios}, the $p_{\rm T}$ dependence of $\psi$(2S)-to-J/$\psi$ ratio in Pb--Pb collisions is compared with the corresponding ratio in pp collisions. The $\psi$(2S)-to-J/$\psi$ ratio increases as a function of $p_{\rm T}$ in both Pb--Pb and pp collisions, with a milder rise in the former case. The corresponding double ratio shown in the bottom panel, indicates a significant relative suppression of $\psi$(2S) in Pb--Pb with respect to pp, with no strong $p_{\rm T}$ dependence and reaching a value of $\sim$ 0.5 at high $p_{\rm T}$.

\begin{figure}[ht]
\centering
\includegraphics[scale=0.37]{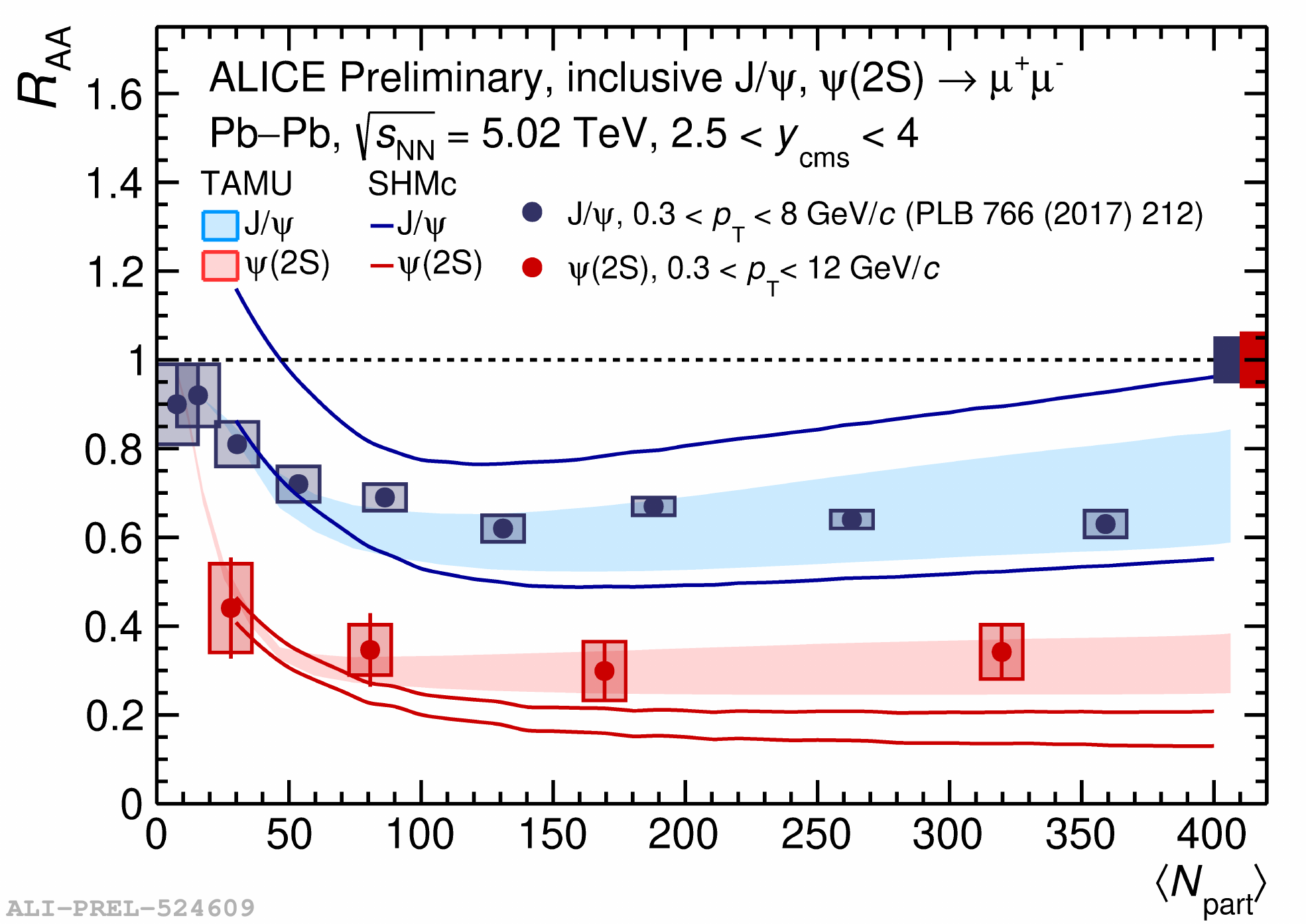}
\includegraphics[scale=0.37]{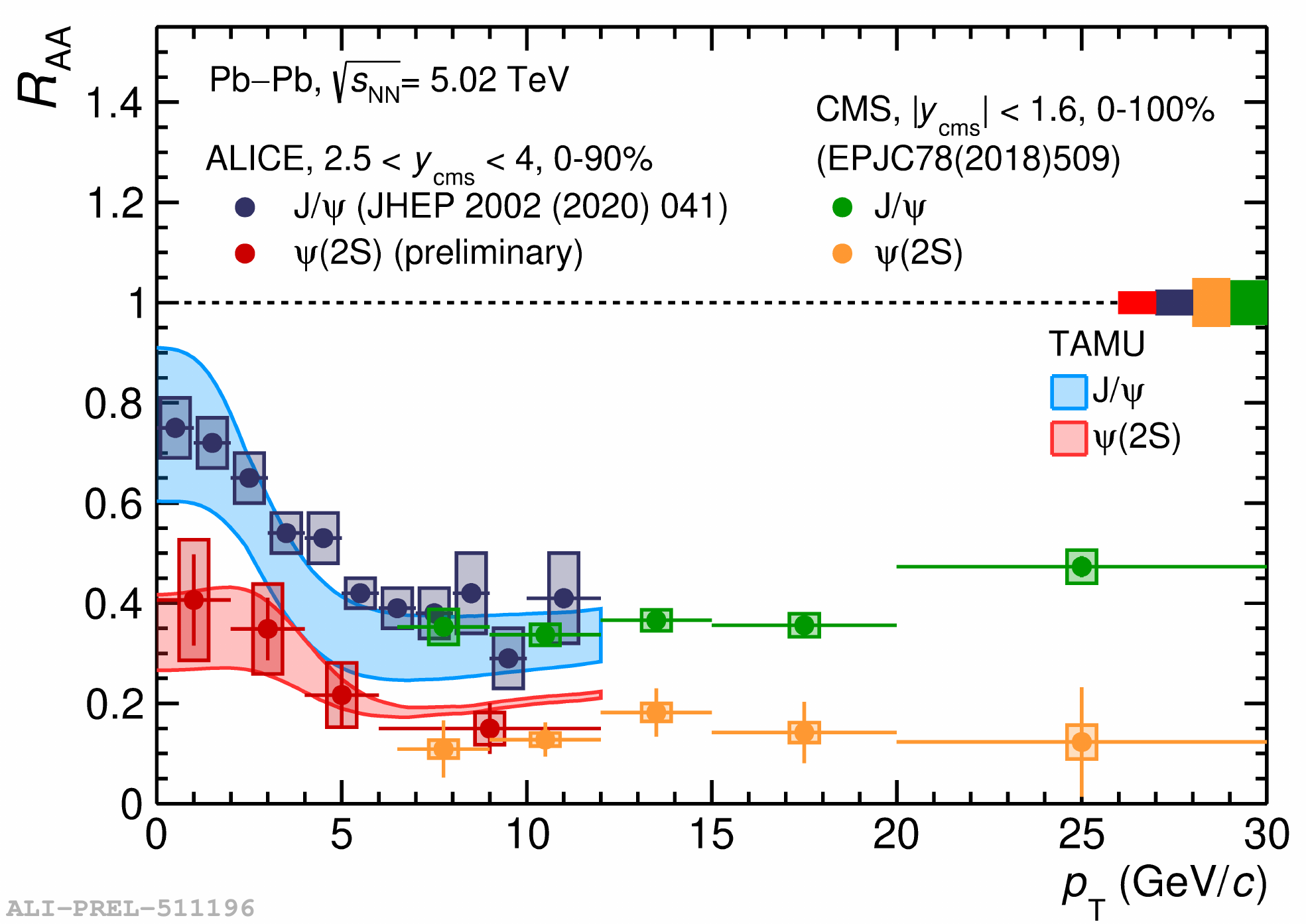}
\vspace{-0.2cm}
\caption{\label{Psi2SJPsiRAA}The $R_{\rm AA}$ of $\psi$(2S) and J/$\psi$ as a function of $\langle N_{\rm part}\rangle$ (left) and $p_{\rm T}$ (right). In the right panel, the ALICE data are compared with CMS results for $|y|$ $<$ 1.6, 6.5 $<$ $p_{\rm T}$ $<$ 30 GeV/$c$ and centrality 0--100 \%~\cite{CMS}. The results are also compared with theoretical predictions from TAMU~\cite{TAMU} (left and right plots) and SHMc~\cite{SHMc,SHMc2} (left plot).}
\end{figure}

Figure~\ref{Psi2SJPsiRAA} shows the nuclear modification factor $R_{\rm AA}$ of J/$\psi$ and $\psi$(2S) measured by the ALICE collaboration as a function of $\langle N_{\rm part}\rangle$ (left panel) and $p_{\rm T}$ (right panel). The results show that $\psi$(2S) is strongly suppressed than J/$\psi$ both as a function of $p_{\rm T}$ and centrality. Flat centrality dependence of $\psi$(2S) $R_{\rm AA}$ is consistent with an $R_{\rm AA}$ value of about 0.4. TAMU model~\cite{TAMU} reproduces the centrality dependence of $R_{\rm AA}$ for both J/$\psi$ and $\psi$(2S), while SHMc~\cite{SHMc,SHMc2} reproduces the J/$\psi$ result but tends to underestimate the $\psi$(2S) $R_{\rm AA}$ in central and semi-central collisions.

The $p_{\rm T}$ dependence of $R_{\rm AA}$ shows a stronger suppression at high $p_{\rm T}$ and increasing trend of $R_{\rm AA}$ towards low $p_{\rm T}$ for both charmonium states. This is a hint of charmonium regeneration. The result is in good agreement with CMS measurements in $|y|$ $<$ 1.6, 6.5 $<$ $p_{\rm T}$ $<$ 30 GeV/$c$ and centrality 0--100 \% at high $p_{\rm T}$~\cite{CMS}. TAMU~\cite{TAMU} model reproduces the $p_{\rm T}$ dependence of $R_{\rm AA}$ for both J/$\psi$ and $\psi$(2S), as it was the case for the centrality dependence.

\section{Summary}
$\psi$(2S) cross section and $\psi$(2S)-to-J/$\psi$ ratio have been measured at $\sqrt{s}$ = 5.02 TeV, with significantly improved precision compared to earlier publication~\cite{Psi2S_Old}. The first accurate measurement of the $\psi$(2S) production in Pb--Pb collisions at $\sqrt{s_{\rm NN}}$ = 5.02 TeV down to zero $p_{\rm T}$ has been reported by ALICE at forward rapidity. $\psi$(2S)-to-J/$\psi$ signle and double ratios, and nuclear modification factor of $\psi$(2S) have been measured. The $\psi$(2S) is more suppressed than the J/$\psi$ as a function of $p_{\rm T}$ and centrality. From the double ratio, a relative suppression by a factor $\sim$ 2 of the $\psi$(2S) with respect to the J/$\psi$ is observed, with almost flat $p_{\rm T}$ or centrality dependence within the uncertainties. The double ratio measurements from NA50 show a more pronounced centrality dependence compared to ALICE. Flat centrality dependence of $\psi$(2S) $R_{\rm AA}$ with values around $R_{\rm AA}$ $\sim$ 0.4 is observed. Increasing trend of the $p_{\rm T}$ dependence of $R_{\rm AA}$ towards low $p_{\rm T}$ both for J/$\psi$ and $\psi$(2S) is a hint of charmonium regeneration. Transport model (TAMU), which includes recombination of charm quarks in the QGP phase, reproduces the $\psi$(2S) $R_{\rm AA}$ and $\psi$(2S)-to-J/$\psi$ ratio better than SHMc model for central events. 

A significant increase of statistical precision is expected in Run 3 and 4 with $L_{\rm int}$ $\sim$ 10 ${\rm nb}^{-1}$. The Muon Forward Tracker (MFT) will allow to separate the prompt charmonium from the contribution originating from beauty hadron decays at forward rapidity in Run 3 and 4.


\begin{thebibliography}{9}

{\bibitem{satz} {H. Satz, J. Phys. G {\bf 32} (2006), R25.}}
{\bibitem{satz2} {H. Satz, Int. J. Mod. Phys. A {\bf 28} (2013), 1330043.}}
\bibitem{reg_QGP} R. L. Thews et al., Phys. Rev. C {\bf 63} (2001), 054905.
\bibitem{reg_phase} P. Braun-Munzinger et al., Phys. Lett. B {\bf 490} (2000), 196.
\bibitem{reg_phase2} A. Andronic et al., Jour. of Phys. G {\bf 38} (2011), 124081.

\bibitem{alice_pp} S. Acharya et al. [ALICE Collaboration], [arXiv:2109.15240 [nucl-ex]].
\bibitem{bpaul} B. Paul et al., J. Phys. G: Nucl. Part. Phys. {\bf 42} (2015), 065101.

\bibitem{alice} K. Aamodt et al. [ALICE Collaboration], JINST {\bf 3} (2008), S08002.

\bibitem{Psi2S_Old} S. Acharya et al. [ALICE Collaboration], Eur. Phys. J. C {\bf 79} (2019), 402.

\bibitem{MCacci} M. Cacciari, S. Frixione, N. Houdeau, M. L. Mangano, P. Nason, and G. Ridolfi, JHEP {\bf 10} (2012) 137.
\bibitem{But} M. Butensch$\ddot{\rm o}$n and B. A. Kniehl, Phys. Rev. Lett. {\bf 106} (2011) 022003.
\bibitem{QMa} Y.-Q. Ma, K. Wang, and K.-T. Chao, Phys. Rev. Lett. {\bf 106} (2011), 042002.
\bibitem{QMa2} Y.-Q. Ma and R. Venugopalan, Phys. Rev. Lett. {\bf 113} (2014), 192301.
\bibitem{Cheung} V. Cheung and R. Vogt, Phys. Rev. D {\bf 98} (2018), 114029.

\bibitem{NA50} B. Alessandro et al. [NA50 Collaboration], Eur. Phys. J. C {\bf 49} (2007), 559--567.

\bibitem{TAMU} X. Du and R. Rapp, Nucl. Phys. A {\bf 943} (2015), 147.
\bibitem{SHMc} A. Andronic et al., Phys. Lett. B {\bf 797} (2019), 134836.
\bibitem{SHMc2} A. Andronic et al., Nature {\bf 561} (2018), 321--330.

\bibitem{CMS} A. M. Sirunyan et al. [CMS Collaboration], Eur. Phys. J. C {\bf 78} (2018), 509.

\end{thebibliography}
\end{document}